\documentclass[aps,prb,preprint,superscriptaddress]{revtex4-1}

\usepackage{graphicx}
\usepackage{bm}
\usepackage{amsmath}
\usepackage{amssymb}
\usepackage{hyperref}
\usepackage[version=4]{mhchem}
\usepackage{color}
\usepackage{upgreek}

\begin{document}

\title{Machine-learning identified molecular fragments responsible for infrared emission features of polycyclic aromatic hydrocarbons}

\author{Zhisen Meng}
\affiliation{Laboratory for Relativistic Astrophysics, Department of Physics, Guangxi University, Nanning 530004, China}

\author{Yong Zhang}
\affiliation{School of Physics and Astronomy, Sun Yat-sen University, 519082 Zhuhai, China}

\author{Enwei Liang}
\affiliation{Laboratory for Relativistic Astrophysics, Department of Physics, Guangxi University, Nanning 530004, China}

\author{Zhao Wang}
\email{zw@gxu.edu.cn}
\affiliation{Laboratory for Relativistic Astrophysics, Department of Physics, Guangxi University, Nanning 530004, China}

\begin{abstract}

Machine learning feature importance calculations are used to determine the molecular substructures that are responsible for mid and far-infrared (IR) emission features of neutral polycyclic aromatic hydrocarbons (PAHs). Using the extended-connectivity fingerprint as a descriptor of chemical structure, a random forest model is trained on the spectra of $14,124$ PAHs to evaluate the importance of $10,632$ molecular fragments for each band within the range of $2.761$ to $1172.745\,\upmu$m. The accuracy of the results is confirmed by comparing them with previously studied unidentified infrared emission (UIE) bands. The results are summarized in two tables available as Supplementary Data, which can be used as a reference for assessing possible UIE carriers. We demonstrate that the tables can be used to explore the relation between the PAH structure and the spectra by discussing about the IR features of nitrogen-containing PAHs and super-hydrogenated PAHs.

\end{abstract}


\maketitle

\section{Introduction}
A series of unique interstellar mid-infrared (IR) spectral emission features are widely detected in the Milky Way and extragalactic galaxies. Their main bands appear around $3.3$, $6.2$, $7.7$, $8.6$, $11.2$, and $12.7\,\upmu$m are called unidentified infrared emission (UIE) features. It has been recognized since early years that UIE bands could stem from the vibration of \ce{C-H} and \ce{C-C} chemical bonds in organic molecules. This leads to the widely-accepted hypothesis that the UIE carriers could be polycyclic aromatic hydrocarbons (PAHs; \citealt{Leger1984,Allamandola1985}).  In view of the important role of PAHs in the evolution of interstellar medium (ISM), many efforts are devoted to identifying the PAH molecular structures from UIE bands \citep{Peeters2011}. Despite a few PAHs have likely been detected via radio telescopes in recently years, {including PAHs containing \ce{C-N} bonds} \citep{Mcguire2018,Mcguire2021,Burkhardt2021}, the identification of interstellar PAHs as UIE carriers still remain challenging, due to the vast variation of the PAH chemical structures and the complex structure-emission relation \citep{Li2020}.

The IR emission features of PAHs are ascribed to characteristic vibrational modes of bonds or functional groups of specific types \citep{Peeters2021}. Traditionally, the $3.3\,\upmu$m feature is mainly due to \ce{C-H} stretching, the $6.2\,\upmu$m feature to \ce{C-C} stretching, the $7.7\,\upmu$m feature to coupled \ce{C-C} stretching and \ce{C-H} in-plane bending, the $8.6\,\upmu$m feature to the \ce{C-H} in-plane bending and the $10-15\,\upmu$m features to the \ce{C-H} out-of-plane bending, etc. {However, previous studies indicate that observed spectra cannot be adequately explained without considering the specific atomic surroundings of the chemical bond.} For instance, the shift of IR peaks around $3.3\,\upmu$m was not well explained, until the local atomic environment surrounding the \ce{C-H} bond is taken into account \citep{Bauschlicher2009}. The mixed emission features between $10$ and $15\,\upmu$m is found to rely on the configuration of adjacent aromatic rings carrying different numbers of \ce{C-H} bonds \citep{Hony2001}. {The subsequent investigations concentrated on specific subclass of PAHs as motivated by their potential existence in ISM, accounting for ring defects \citep{Ricca2011,Yu2012,Ottl2014,Bauschlicher2015,Devi2020}, side groups \citep{Sadjadi2015,Bauschlicher2016,Yang2017,Buragohain2020}, edges \citep{Sandford2013,Candian2014,Ricca2018,Ricca2019,Yang2020}, charge \citep{Cecchi-Pestellini2008,Devi2022,Maragkoudakis2020}, size \citep{Draine2007,Bauschlicher2008,Bauschlicher2009,Andrews2015}, doping of heteroatoms \citep{Buragohain2015,Canelo2018,Oliveira2021}, etc. Despite these efforts, a comprehensive understanding of the correlation between the structure and spectrum of PAHs  across the broad IR region is still missing.} 

The advent of the big data era has brought a new opportunity to improve the situation. Spectra of a large number of organic compounds are made available in database systems. Remarkably, the NASA Ames PAH IR Spectroscopic Database (PAHdb) has compiled over four thousand spectra of PAHs using ab-initio calculations {\citep{Bauschlicher2018nasa}} and experiments \citep{Mattioda2020nasa}. This high-quality data has inspired the use of machine learning (ML) methods for a more comprehensive view of the complex spectrum-structure relationship \citep{Mcgill2021,Laurens2021,Calvo2021}. Recently, \citet{Kovacs2020} trained a neural network (NN) model for predicting IR spectra of PAHs directly from their chemical structures, and demonstrated excellent predictive skill of the ML model for out-of-sample inputs. In this study, we built a ML model to explore the complex relation between the subtle molecular structure of PAHs and their IR emission features. The molecular structural origins have been identified for individual bands in a wide spectral region. 

\section{Methods}

\subsection{Input: Molecular descriptor}

The ML prediction of physical properties of a molecule strongly depends on how the molecular chemical structure is mathematically described. Traditionally, molecular geometry in chemical spectra databases is represented using cartesian coordinates. However, cartesian coordinates are not a suitable molecular descriptor for ML since they are not invariant to translation, rotation, and permutation of atom numbers. Therefore, in ML, the cartesian coordinates is often transformed into an invariant molecular descriptor such as the extended-connectivity fingerprint (ECFP; \citealt{Rogers2010}), the distance or coulomb matrix \citep{Ivanciuc2000,Rupp2012}, or so on. In particular, the ECFP shows superior predictive power for the ML prediction of PAH IR spectra compared to other invariant descriptors \citep{Meng2021}.  

\begin{figure}
\centering
\includegraphics[width=\columnwidth]{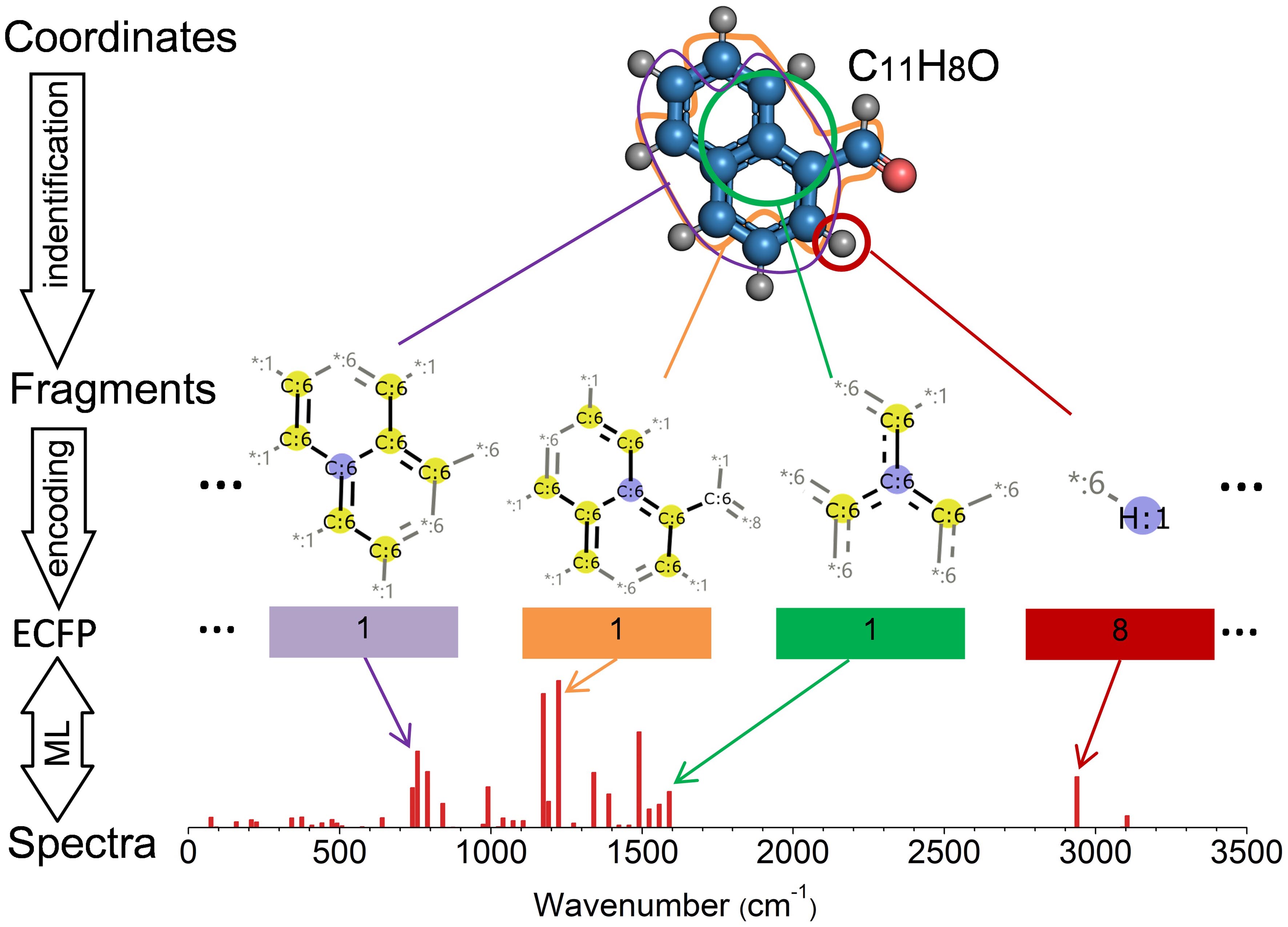}
\caption{Schematic of the processes for generating the ECFP of 1-Naphthaldehyde. Arrows in colors indicate the correlation between molecular fragments and IR bands. {The purple circle corresponds to the fragment center, the yellow circle represents aromatic atoms, and the asterisk symbol represents adjacent atoms. The atomic number is indicated by the digit following the colon. The gray chemical bonds depict the adjacent portion of the fragments.}}
\label{F1}
\end{figure}

The ECFP encodes the local neighborhoods and bonding connectivity of each atom in a molecule into a series of digital identifiers, each of which represents an individual substructure of the molecule, namely a molecular fragment. To identify possible fragments in a molecule, we used the algorithm of \citet{Morgan1965} that systematically records the neighborhood of each non-hydrogen atom into multiple circular layers up to a cutoff radius ($R\rm _{cut}$), as implemented in RDKit\footnote{\url{http://www.rdkit.org/}\label{web_rdkit}}. Thereby, an integer identifier (a digit) is created for each molecular fragment to count how many times it repeatedly appears at different locations in the molecule. For example, as shown in Figure \ref{F1}, eight \ce{C-H} bonds can be found in \ce{C11H8O}. So, the number $8$ is assigned to the identifier of \ce{C-H} in the ECFP of \ce{C11H8O}. Every digit like this representing a specific fragment, is below called an \textit{ECFP feature} (to distinguish from the term `emission feature' for the IR spectrum) in ML. The ECFP is finally created by collecting all the ECFP features into a digital string. The selection of the $R \rm_{cut}$ cutoff radius is a compromise between predictive accuracy and the size of the fragment library, since the prediction error reduces and the fragment number increases with increasing $R \rm_{cut}$ (see the Section 1 of the Supplementary Information). Using this approach, $10,632$ molecular fragments were in total identified for the PAH molecules considered in this work with $R\rm _{cut}=3$.

\subsection{Output: IR spectra}

\begin{figure}
\centering
\includegraphics[width=\columnwidth]{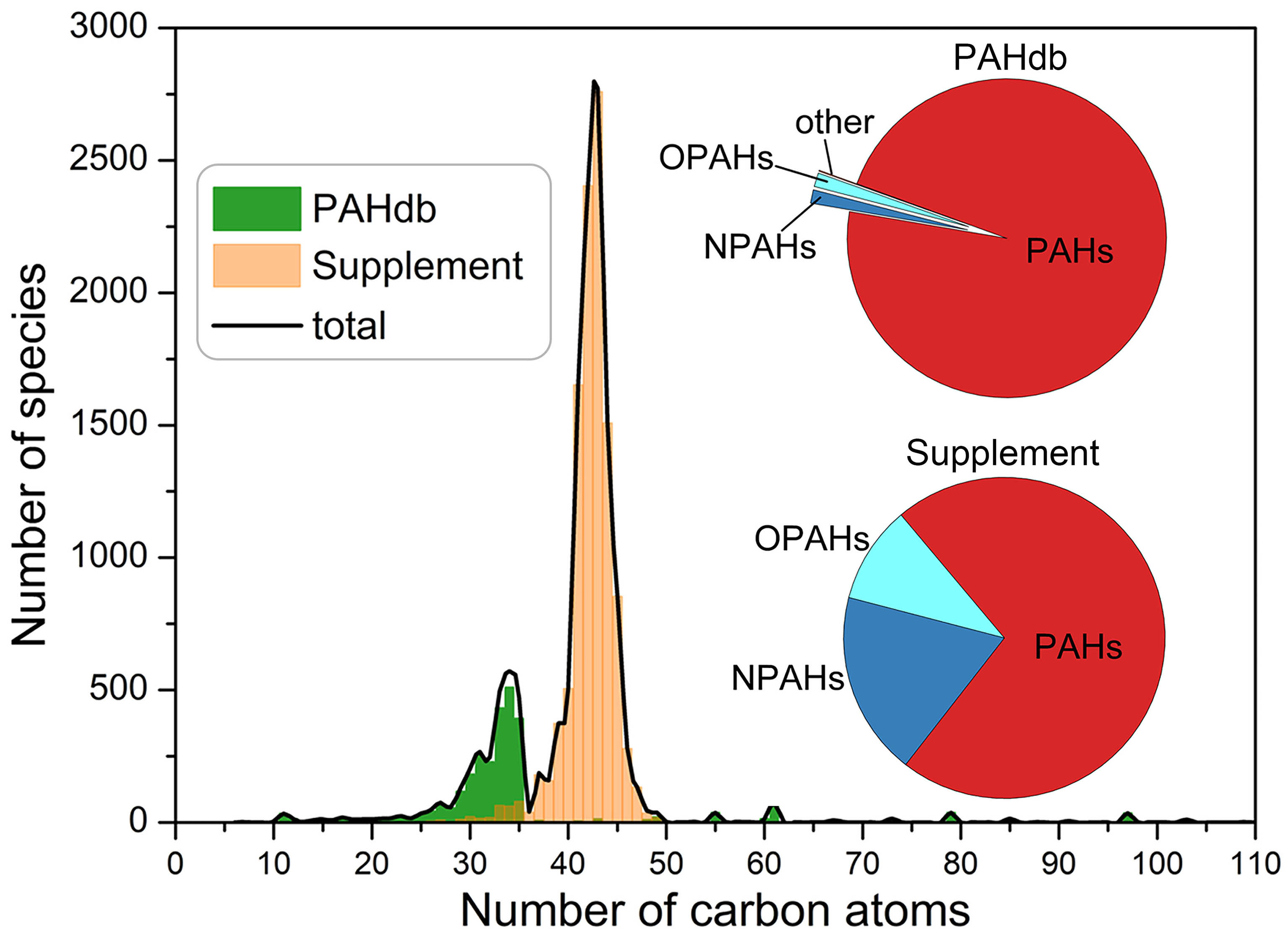}
\caption{Comparison of PAH size distributions in the two datasets (green for PAHdb, orange for Supplement). Insets show the proportion of PAHs of different types: `PAHs' stands for molecules with solely C and H, `NPAHs' and `OPAHs' for N- and O- containing ones, respectively.}
\label{F2}
\end{figure}

The ECFPs and the IR spectra of $14,124$ PAHs were used as the input and the output for training the ML model, respectively. The output dataset includes the theoretical spectral diagrams of $2,925$ neutral PAHs from the 3.2 version of the PAHdb\footnote{\url{https://www.astrochemistry.org/pahdb/}}, after excluding $1,213$ charged PAHs and $95$ ones whose cartesian coordinates cannot be successfully transformed into ECFPs by RDKit. We here focused exclusively on neutral PAHs because the topological descriptor we used was unable to distinguish effectively between neutral PAHs and ions, despite the availability of spectra for charged PAHs in the PAHdb dataset. As a result, our model was not very accurate in predicting ions, despite our attempts to modify the ECFP. The majority of PAHs in PAHdb contains less than $35$ C (see Figure~\ref{F2}). However, previous studies indicate that UIE bands could more possibly come from larger PAHs (e.g. with $\sim 50$ C) \citep{Tielens2008}. To complement PAHdb, we have computed the IR spectra of $11,199$ PAHs with larger size as an additional dataset (namely Supplement, see distribution in Figure~\ref{F2}). {A significant distinction between the molecules present in PAHdb and the supplementary dataset lies in their disparate size distributions, resulting in variations in the considered fragments. For example, PAHdb yielded 6,468 fragments in total (with $R\rm _{cut}=3$), whereas the inclusion of the supplementary dataset resulted in the discovery of 10,632 fragments.} These spectra have been computed using a Gaussian 16 B.01 \citep{Frisch2016} implementation of density functional theory (DFT) at the B3LYP/6-31G level {with a scaling factor of $0.961$}, one of the approaches used by PAHdb \citep{Boersma2014nasa,Bauschlicher2018nasa}. 

It should be noted that this work includes a selection of unsaturated PAHs to ensure the representation of the diverse range of potential UIE carriers, taking into account the presence of radicals in the interstellar molecules that have been detected previously \citep{Guelin2022,McGuire2022}. The Supplementary Data III provides the chemical formulas, number of unpaired valence electrons, spin multiplicities, SMILES strings, and xyz coordinates of the PAHs that carry the additional spectra. We also note that in this study, the dataset was computed using the harmonic assumption same as PAHdb, which was necessary due to the large size of the systems, even though anharmonic calculations can provide better agreement with experimental data \citep{Mackie2015the, Mackie2016, Lemmens2019}.

{The discrete IR intensity is converted to a histogram with a bin width of 16.65 cm$^{-1}$, determined through the utilization of Knuth's Bayesian rule as implemented in the Astropy package \citep{Knuth2006}.} This results in $171$ bands in the spectral region between $2.761$ and $1172.745\,\upmu$m ($8.527$ to $3620.590$\,cm$^{-1}$). Note that there are $54$ intervals with no band at all, most of which are in the ranges of $3.720-5.030\,\upmu$m ($1989.340-2688.450$\,cm$^{-1}$) and of $2.761-3.120\,\upmu$m ($3204.450-3620.590$\,cm$^{-1}$). Bands beyond $2.761\,\upmu$m ($3620.590$\,cm$^{-1}$) were discarded as there is only a single compound in the whole database contributing to this region. 

Using the method introduced in the previous subsection, we identified $10,632$ ECFP fragments from the molecular structures of the $14,124$ PAH molecules, which were used to compose the ECFPs as the input for training the ML model. The input and output datasets, along with the code of ML program, are provided as Supplementary Data IV for the ease of the readers who are interested in repeating the calculations.

\subsection{ML algorithm}
\label{sec:ML}

The random forests (RF) algorithm was used to train the model as implemented in the open-source Scikit-Learn library \citep{Pedregosa2011}. The RF algorithm is chosen because it has a built-in metric for the feature importance that is practical for the task of this work, in addition to its efficiency and natural resistance to overfitting \citep{Svetnik2003}. A RF is a combination of decision trees $T$, every of which comprises a number of nodes $t$ connected through splitting branches. The hyperparameters (such as the number of trees, etc) were previously adjusted for predicting PAH IR spectra \citep{Kovacs2020,Meng2021}, as provided within the ML code in Supplementary Data IV. 

In Section 2 of the Supplementary Information, we conducted a test to assess the impact of changes in hyperparameters and datasets on our model outputs. These results indicate that our model remains resilient to such variations. Details regarding the accuracy of the model, which includes the outcomes of a 5-fold cross-validation, are available in Section 3 of the Supplementary Information. These results demonstrate that our models exhibit strong out-of-sample predictability.

For each of the $171$ frequency intervals, we trained a RF model to assess the importance of each ECFP feature on a specific IR emission feature, since an IR emission band could strongly be related to a specific molecular fragment that are represented by an ECFP feature $X_{m}$ (e.g. indicated by the arrows in Figure \ref{F1}). In ML analysis, this correlation is evaluated by the importance of $X_{m}$ for predicting the spectrum, namely the feature importance $FI (X_{m})$. It can be calculated by adding up the weighted impurity decreases $p(t)\Delta i(s_t, t)$ for all nodes where $X_m$ is used, averaging over all $N_T$ trees in the forest,

\begin{center}
\begin{equation}
\label{eq1}
FI\left(X_{m}\right)=\frac{1}{N_{T}} \sum_{T} \sum_{t \in T: v\left(s_{t}\right)=X_{m}} p(t) \Delta i\left(s_{t}, t\right),
\end{equation}
\end{center}

\noindent where $p(t)$ is the proportion $N_t/N$ of samples reaching $t$, $v(s_t)$ is the variable used in the split $s_t$. The Gini index is used as the impurity function $i(t)$ \citep{Louppe2013}. The sum of the importance of all ECFP features is normalized to $1.0$. As $10,632$ ECFP features are included in this work, the average value of $FI\left(X_{m}\right)$ will be as low as $9.4\times10^{-5}$. A $FI$ at the order of $>10^{-3}$ or $10^{-2}$ could therefore be prominent over a band. We note that there is no definite correlation between $FI$ and the band’s intensity. {The random forest algorithm incorporates a random component that may introduce uncertainty (noise) into the outcomes. After conducting a preliminary assessment, it has been determined that the average random noise in our feature importance results is estimated to be around $3.58$\% of $FI(Xm)$.}

\section{Results and Discussion}

The importance of an ECFP feature, denoted as $FI$, indicates the level of importance of a molecular fragment in relation to a given emission band. By evaluating $FI$, we have identified the molecular fragments responsible for each band between $2.761$ and $1172.745\,\upmu$m. The findings are summarized in two oversized tables, which are included as the Supplementary Data I and II of this article. The first table lists the top 10 molecular fragments in descending order of importance for each emission band, while the second table includes the top 100. These tables offer a valuable tool for assessing potential carriers of UIE. It is important to note that while $FI$ aids in identifying the molecular structure class associated with a specific band, there is no definitive correlation between $FI$ and the absolute band intensity. Considering the extensive amount of information encompassed by these tables, we have carefully selected a few emission features that are of significant astronomical interest for detailed discussion in the subsequent text.

\begin{figure}
\centering
\includegraphics[width=\columnwidth]{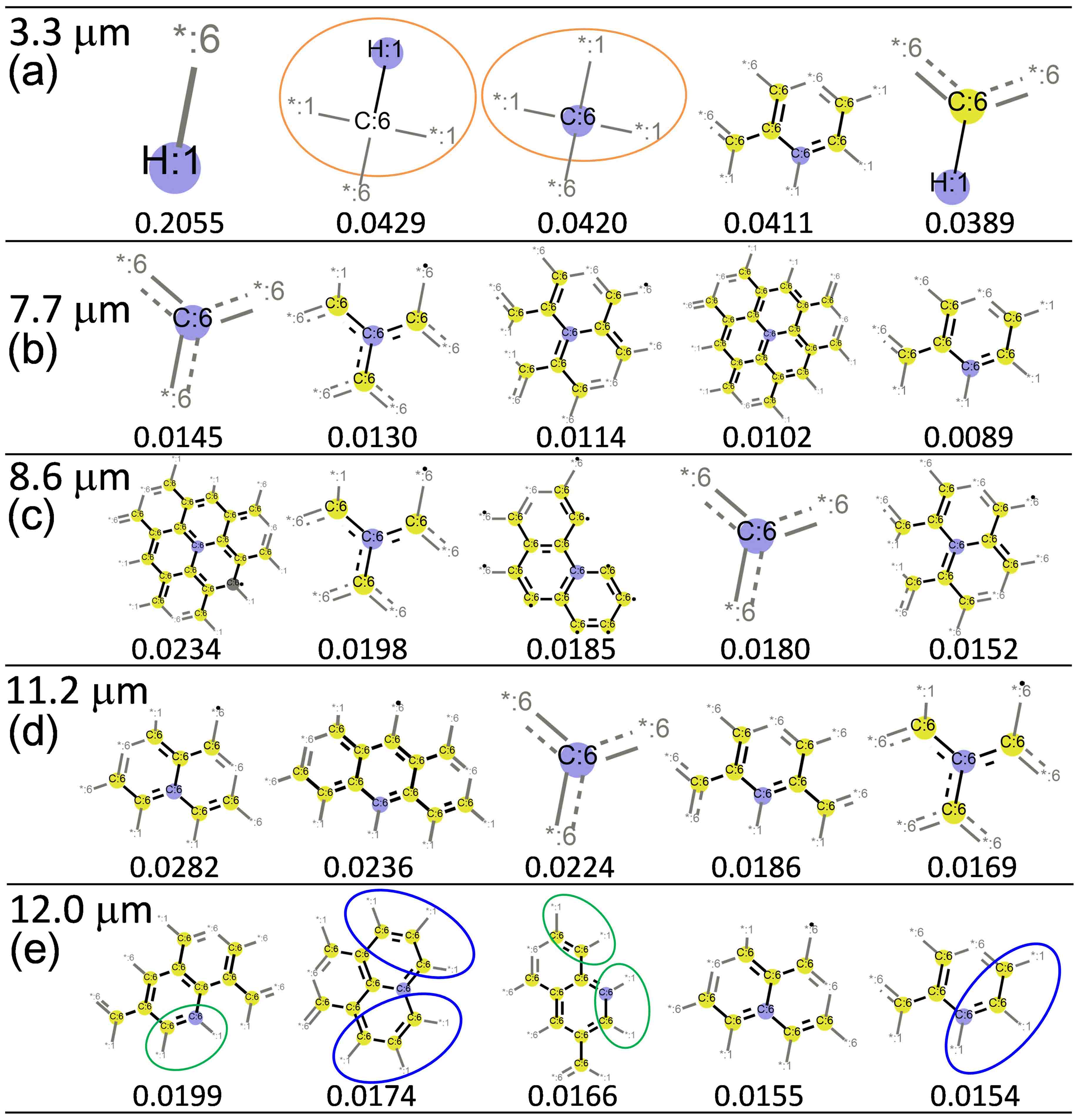}
\caption{Five most important fragments for the bands at $3.3$, $7.7$, $8.6$, $11.2$, and $12.0\,\upmu$m, sorted from left to right with descending $FI$ (printed at the bottom).}
\label{F3}
\end{figure}

To validate our model, the results are first compared with the established sources of extensively studied UIE bands. These include the $3.3$, $7.7$, $8.6$, $11.2$, and $12.0\,\upmu$m bands that are known to arise from \ce{C-H} stretching, \ce{C-C} stretching, \ce{C-H} in-plane bending, and \ce{C-H} out-of-plane bending vibration, respectively. {It is important to note that the results presented for the 7.7, 8.6, and 12.0 $\upmu$m bands in this study pertain to neutral PAHs, despite these bands being predominantly associated with strong cationic components \citep{Bauschlicher2009}.} The ML results shown in Figure~\ref{F3} are found to be consistent with the classical theory. Interestingly, the methyl group is found to make a noticeable contribution to the band at $3.3\,\upmu$m (as highlighted by the orange ellipses in Figure~\ref{F3}), in addition to its well-known contribution to the $3.4\,\upmu$m band \citep{Steglich2013}. Contrary to the results of \citet{Candian2012} suggesting that UIE bands slightly blue-shifted from $3.3\,\upmu$m may originate from PAH with bay-like structures, the molecular fragments with bay-like structures were not found to dominate the bands at $3.29\,\upmu$m.

Regarding the $7.7\,\upmu$m band, the fragments responsible, as illustrated in Figure~\ref{F3}(b), suggest that \ce{C-C} bonds within the carbon skeleton hold greater significance compared to those located at the edges. {As shown in Panel (d), it is seen that the 11.2 $\upmu$m band is primarily characterized by solo \ce{C-H} bonds (without contiguous H atoms), aligning with the out-of-plane bending vibration mode of neutral PAHs previously investigated by \cite{Maragkoudakis2020}.} In the case of the $12.0\,\upmu$m band, the \ce{C-H} bonds of the trio type (with three contiguous H atoms) are found to be equally prominent as the commonly considered duo-type bonds, as highlighted by the blue and green ellipses in Panel (e). Furthermore, carbon skeleton fragments demonstrate relatively high importance across the entire spectrum of bands. This observation is to be expected, as these fragments are indicative of the characteristic aromatic structure. However, the positioning of these fragments among the top fragments for a given band indicates a reliance on the specific carbon skeleton structure associated with that particular band, as illustrated in the section 4 of the Supplementary Information. {It is noteworthy that a wavelength of $1172.745\,\upmu$m is present, at which emission bands are exhibited by 904 PAHs in our dataset. As demonstrated in Section 5 of the Supplementary Information, a visualization of a sample molecule's vibration suggests that this feature is likely correlated with the out-of-plane twisting of the molecule.}

\begin{figure}
\centering
\includegraphics[width=\columnwidth]{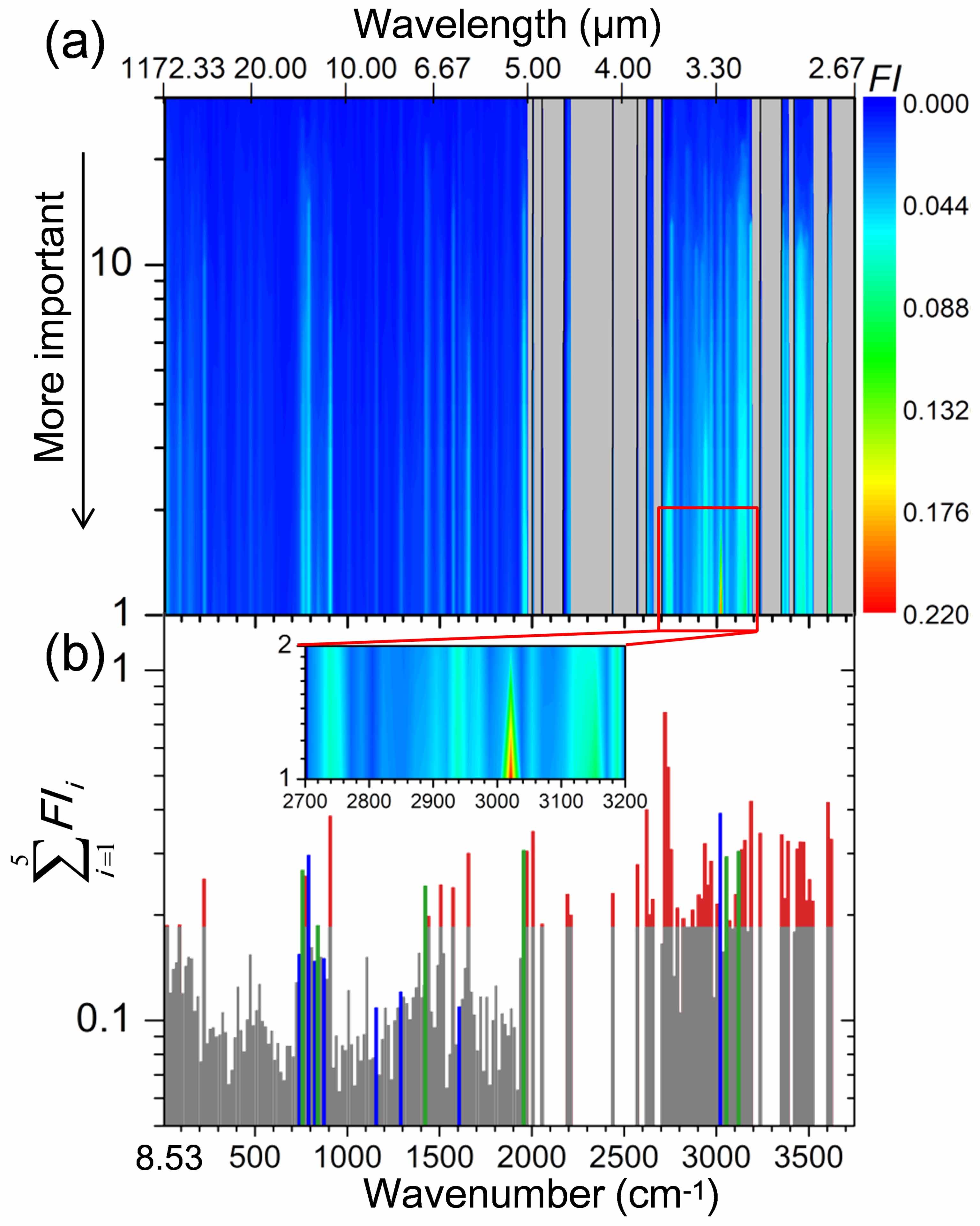}
\caption{(a) $FI$ distribution for the bands between $2.67$ and $1172.33\,\upmu$m. Upwards along the vertical axis, the importance of $100$ ECFP features are sorted from high to low, and is represented by the color scale. The regions with no spectrum are shown in gray. (b) Sum of $FI$ of the five most important fragments for different bands. {The illustration is a zoom of Panel (a) at a range of about $3.3\,\upmu$m (3000 cm$^{-1}$). The $FI$ maximum shown for this band is associated with fragment 1 in Figure 3(a).}}
\label{F4}
\end{figure}

Statistically, when a band is dominated by the vibration characteristics of a particular molecular fragment, the ECFP feature corresponding to which will exhibit an importance significantly higher than that of other fragments. Consequently, the distribution of $FI$ of a distinguishing band should be non-uniform, and may represent a particular structural class of PAHs.  To visualize the $FI$ distribution for different bands, we used a color scale in Figure~\ref{F4}(a). Notably, several IR bands exhibit a non-uniform color scale, suggesting that they may be distinguished by the dominance of some specific molecular fragments.

To display the non-uniform distributions more clearly, we plot the sum of $FI$ of the five most important fragments for different bands in Figure~\ref{F4}(b). Roughly, the sum is supposed to be high for a distinguishing band. The blue bars in this figure highlight a number of prominent bands that are intensively discussed in the literature, such as the aforementioned ones at $3.3$, $6.2$, $7.7$, $8.6$, $11-15\,\upmu$m. The $3.21$, $3.27$, $5.1$, $7.0$, $11.9$, $13.2\,\upmu$m bands (green bars) are also noticeable in UIE \citep{Bauschlicher2009,Bauschlicher2013,Sundararajan2018,Ricca2019}. Our findings also identify several unique bands (indicated by red bars) that may be of interest for future astronomical observations of PAHs (with $FI$ sum above the mean of the blue bars). The wavelengths associated with these bands are $44.46$, $12.92$, $11.02$, $6.64$, $6.36$, $6.04$, $5.07$, $4.99$, $4.98$, $4.86$, $4.1$, $4.56$, $4.52$, $4.1$, and $3.89-2.76\,\upmu$m.

\begin{figure}
\centering
\includegraphics[width=\columnwidth]{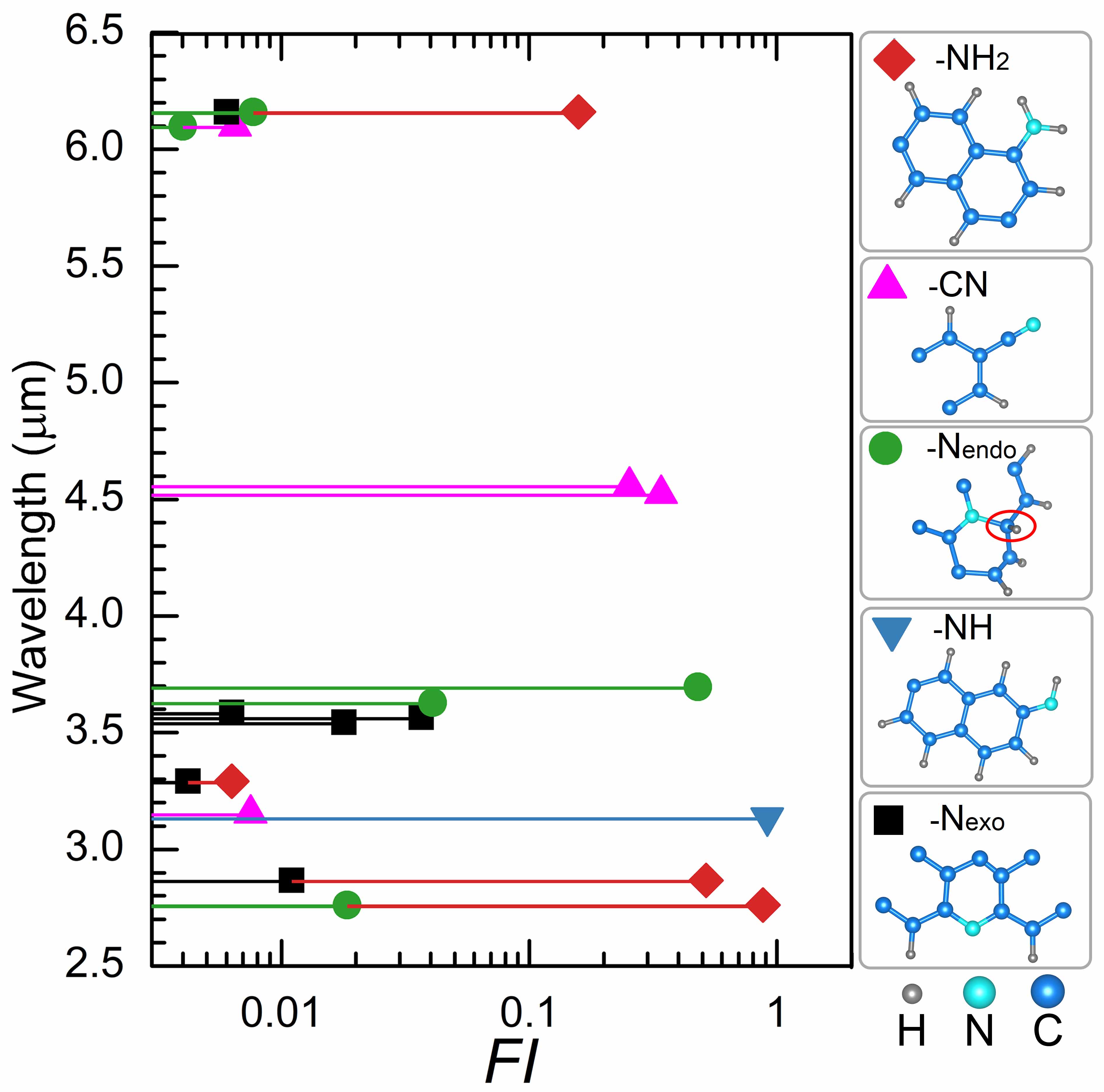}
\caption{Left: $FI$ of fragments in different classes for the bands between $2.5$ and $6.5\,\upmu$m. Right: Example fragments for five NPAH structural classes.}
\label{F5}
\end{figure}

Among the distinguishing emission features shown in Figure~\ref{F5}, the $2.9$, $3.1$, and $3.7\,\upmu$m ones are characteristic for nitrogen-containing PAHs (NPAHs). NPAHs are of importance for astronomical observations aiming at the abiogenesis question as they could be building blocks of biomolecules. Previous studies indicate that the stretching of \ce{N-H} bonds gives rise to emission bands at $2.8-3.1\,\upmu$m \citep{Vats2022}, while bands around $4.5\,\upmu$m are suspected to stem from the stretching of \ce{C#N} bonds \citep{Allamandola2021}. Here, further structural dependence of these emission features is revealed by ML. Our results indicate that the bands at different frequencies in the spectral region between $2.5$ and $6.5\,\upmu$m is respectively dominated by NPAHs of different structural classes, namely amine (\ce{-NH2}), imine (\ce{-NH}), cyanide (\ce{-CN}), N embedded inside (N$\rm _{endo}$), and on edges (N$\rm _{exo}$), as shown in Figure~\ref{F5}. We see that the bands at $2.761/2.867$, $3.136$, and $4.523/4.559\,\upmu$m are dominated by the fragments of the \ce{-NH2}, \ce{-NH}, and \ce{-CN} types, respectively. This is in line with with previous studies that have attributed the $2.8-3.1\,\upmu$m bands to the stretching of \ce{N-H} bonds and the $4.5\,\upmu$m bands to the stretching of \ce{C#N} bonds. For instance, \citet{Rozenberg2021} reported that the $2.801$ and $2.897\,\upmu$m bands correspond to the antisymmetric and symmetric stretching of \ce{N-H} bonds in \ce{-NH2}, respectively, while \citet{Gobi2019} suggested that the $3.1\,\upmu$m band is due to the stretching of the \ce{N-H} bond in \ce{-NH}. Our results also identify other emission features that could be attributed to NPAHs of different structural types, such as amine, imine, and cyanide, highlighting their potential significance for future astronomical observations. 

Specifically, our results suggest that NPAHs containing N$\rm _{endo}$ and \ce{-NH2} fragments are characterized by distinctive emission peaks at $3.674$ and $6.160\,\upmu$m, respectively. The blue-shift of the band at $6.2\,\upmu$m (known as the class A band) has been previously proposed as a tracer for interstellar NPAHs, with N$\rm _{endo}$-type fragments responsible for the blue-shifted emission \citep{Hudgins2005,Ricca2021}. However, our results indicate that the blue-shifted emission at $6.16\,\upmu$m is mainly attributed to \ce{-NH2}-type fragments. Figure~\ref{F5} reveals that fragments with the \ce{-NH2} sidegroup are ten times more important than N$\rm _{endo}$ or N$\rm _{exo}$-type fragments in contributing to this emission feature. Additionally, we observe that the emission at $3.674\,\upmu$m is highly correlated with N$\rm _{endo}$ fragments, specifically those with $sp^{3}$-hybridized C, such as the one marked with a red circle in the right panel of Figure~\ref{F5}. These results shed new light on the structural dependence of NPAH emission features. 

Accompanying the dominant $3.3\,\upmu$m emission feature, emissions near $3.4\,\upmu$m is widely detected toward interstellar regions \citep{Sloan1997}. Their origin is under debate between several paradigms, including bond anharmonicity \citep{Barker1987}, aliphatic side-groups \citep{Joblin1996} and super-hydrogenation (SH) \citep{Schutte1993}. Recently, \citet{Kwok2022} show that a number of observations contradict the anharmonicity explanation \citep{Van_Diedenhoven2004,Goto2003}. Also, \citet{Sandford1991} found that it is hard to fit the relative intensity of the bands at $3.40$, $3.46$, $3.51$, and $3.56\,\upmu$m towards NGC 7027 using the spectra of PAHs with aliphatic side-groups. On the other hand, evidences have emerged in recent years supporting the SH origin of IR spectra \citep{Materese2017,Mackie2018,Yang2020}. However, it remains challenging to systematically match the relative intensity of the emissions around in the $3.4\,\upmu$m range.

\begin{figure}
\centering
\includegraphics[width=\columnwidth]{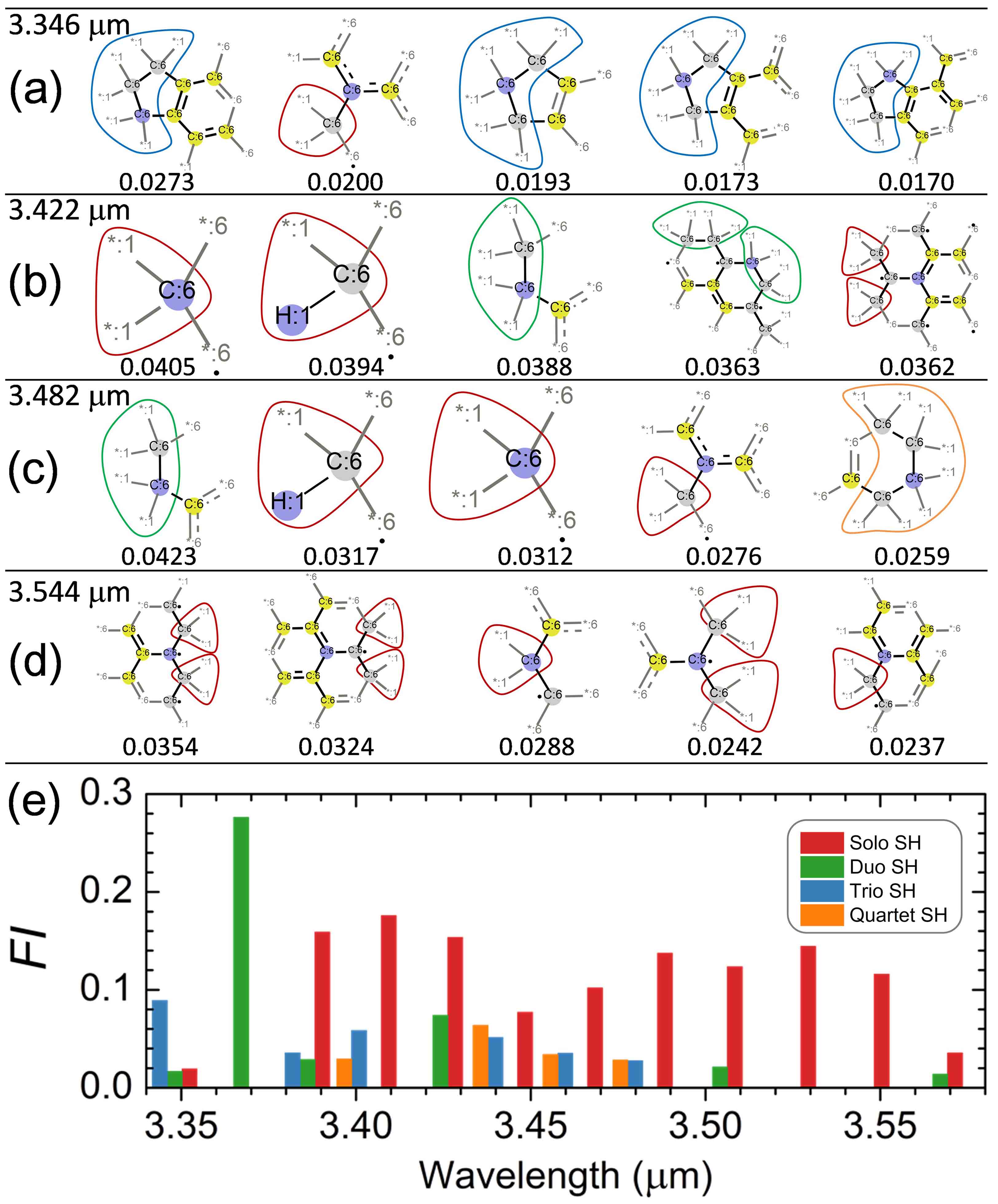}
\caption{Five most important fragments for the bands at $3.346$ (a), $3.422$ (b), $3.482$ (c) and $3.544\,\upmu$m (d), sorted from left to right with descending importance. (e) Distribution of the $FI$ of different SH substructure classes for the bands near $3.4\,\upmu$m.}
\label{F6}
\end{figure}

Moreover, our ML results reveal a dependence of the emission features around $3.4\,\upmu$m on different substructures of SH PAHs. As shown in Figure~\ref{F6}(a)$-$(d), the bands at $3.346$, $3.422$, $3.482$, and $3.544\,\upmu$m are attributed to SH PAH substructures of four different classes. These classes include SH C sites without adjacent SH atoms (referred to as ``Solo SH", illustrated by red coils), those with two adjacent SH atoms (termed ``Duo SH", represented by green coils), those with three adjacent SH atoms (labeled ``Trio SH", depicted by blue coils), and those with four adjacent SH atoms (called ``Quartet SH", illustrated by orange coils). Figure~\ref{F6}(e) shows that the emission of Solo SH PAHs is red-shifted in comparison to that from SH PAHs belonging to the other three classes. {Furthermore, in the 3.523 $\upmu$m bands, a fragment referred to as ``tertiary C" in prior studies \citep{Pla2020} has been identified. However, its significance is slightly lower compared to superhydrogenated molecular fragments.}

\section{Conclusions}

The analysis of ML feature importance was employed to determine the structural classes of PAHs that could potentially account for IR emission bands across a broad spectral range. The resultant tables, accessible as Supplementary Data in this article, establish a connection between the chemical structures of $14,124$ neutral PAHs and their corresponding IR bands. To validate the findings, several well-known unidentified infrared emission (UIE) bands were utilized as benchmarks. Furthermore, numerous distinctive bands were identified, which hold potential significance for astronomical observations of PAHs, many of which fall within shorter wavelengths compared to the commonly examined bands in the PAH "fingerprint" region. To showcase the potential utility of the Supplementary Data in exploring the intricate relationship between PAH structures and spectra, our results were applied to investigate the origin of emission features in NPAHs and SH PAHs. The obtained results largely align with previous studies but provide additional insights into the structural dependency of bands within the $2.7-6.2\,\upmu$m range. {Beyond the commonly studied molecular substructure types, we have identified distinct molecular substructure types through analyzing the statistics of important fragments across all bands. These include substructures with topological defects (such as C4, C5, and C7 rings), Si substitution, and hydroxyl groups. Furthermore, the substructures characterized by zigzag, armchair, or gulf edges hold potential significance in unraveling the formation processes of PAHs \citep{Qi2018,Hanine2020}. We highlight these structures as suggestions for further investigation.}\\


\section*{Data Availability}

Supplementary Data I and II contain extensive tables that lists the top-10 and the top-100 molecular fragments responsible for the spectral bands between $2.761$ and $1172.745\,\upmu$m, respectively. Supplementary Data III provides the chemical formulas, number of unpaired valence electrons, spin multiplicities, SMILES strings, and xyz coordinates of the PAHs that carry the additional spectra. Supplementary Data IV includes the ML training code. Supplementary Information contains validation results of the ML model. All supplementary materials are available on Zenodo at DOI:\href{https://doi.org/10.5281/zenodo.7758208}{10.5281/zenodo.7758208}. 

%

\end{document}